\documentclass[prl,twocolumn,superscriptaddress,showpacs,floatfix]{revtex4}
\usepackage{graphicx}
\usepackage{verbatim}
\usepackage{hhline}
\usepackage{units}
\usepackage{array}

\usepackage{multirow} \usepackage{rotating}

\newcommand{\ket}[1]{|#1\rangle}

\begin{document}

\title{Realization of Universal Ion Trap Quantum Computation with
  Decoherence Free Qubits}

\date{\today}

\author{T.~Monz}
\affiliation{Institut f\"ur Experimentalphysik, Universit\"at
Innsbruck, Technikerstr. 25, A-6020 Innsbruck, Austria}

\author{K.~Kim}
\altaffiliation{Current Address: Department of Physics and Joint Quantum Institute, University of Maryland, College Park, Maryland, 20742, USA}
\affiliation{Institut f\"ur Experimentalphysik, Universit\"at
Innsbruck, Technikerstr. 25, A-6020 Innsbruck, Austria}

\author{A.~S.~Villar}
\altaffiliation{Current Address: Institute of Optics, Information and Photonics, University of Erlangen-N\"urnberg, Staudtstrasse 7/B2, 91058 Erlangen, Germany}
\affiliation{Institut f\"ur Experimentalphysik, Universit\"at
Innsbruck, Technikerstr. 25, A-6020 Innsbruck, Austria}
\affiliation{Institut f\"ur Quantenoptik und Quanteninformation,
\"Osterreichische Akademie der Wissenschaften, Otto-Hittmair-Platz
1, A-6020 Innsbruck, Austria}

\author{P.~Schindler}
\affiliation{Institut f\"ur Experimentalphysik, Universit\"at
Innsbruck, Technikerstr. 25, A-6020 Innsbruck, Austria}

\author{M.~Chwalla}
\affiliation{Institut f\"ur Experimentalphysik, Universit\"at
Innsbruck, Technikerstr. 25, A-6020 Innsbruck, Austria}

\author{M.~Riebe}
\affiliation{Institut f\"ur Experimentalphysik, Universit\"at
Innsbruck, Technikerstr. 25, A-6020 Innsbruck, Austria}

\author{C.~F.~Roos}
\affiliation{Institut f\"ur Quantenoptik und Quanteninformation,
\"Osterreichische Akademie der Wissenschaften, Otto-Hittmair-Platz
1, A-6020 Innsbruck, Austria}

\author{H.~H\"affner}
\altaffiliation{Current Address: Department of Physics, University of California, Berkeley, 366 LeConte Hall \#7300, CA 94720-7300, USA}
\affiliation{Institut f\"ur Quantenoptik und Quanteninformation,
\"Osterreichische Akademie der Wissenschaften, Otto-Hittmair-Platz
1, A-6020 Innsbruck, Austria}

\author{W.~H\"ansel}
\affiliation{Institut f\"ur Experimentalphysik, Universit\"at
Innsbruck, Technikerstr. 25, A-6020 Innsbruck, Austria}

\author{M.~Hennrich}
\affiliation{Institut f\"ur Experimentalphysik, Universit\"at
Innsbruck, Technikerstr. 25, A-6020 Innsbruck, Austria}
\email{Markus.Hennrich@uibk.ac.at}

\author{R.~Blatt}
\affiliation{Institut f\"ur Experimentalphysik, Universit\"at
Innsbruck, Technikerstr. 25, A-6020 Innsbruck, Austria}
\affiliation{Institut f\"ur Quantenoptik und Quanteninformation,
\"Osterreichische Akademie der Wissenschaften, Otto-Hittmair-Platz
1, A-6020 Innsbruck, Austria}


\begin{abstract}
Any residual coupling of a quantum computer to the environment
results in computational errors. Encoding quantum information in a
so-called decoherence-free subspace provides means to avoid these
errors. Despite tremendous progress in employing this technique to
extend memory storage times by orders of magnitude, computation within
such subspaces has been scarce. Here, we demonstrate the
realization of a universal set of quantum gates acting on
decoherence-free ion qubits. We combine these gates to realize the
first controlled-NOT gate within a decoherence-free, scalable
quantum computer.
\end{abstract}

\maketitle

Decoherence of quantum information can never be completely avoided
even with perfect experimental control. It arises from the coupling of
the quantum system to its environment and eventually limits the
achievable precision of quantum information processing. One method to
tackle faulty information storage is quantum error
correction\cite{Steane_ErrorCorrection,Calderbank_ErrorCorrection,Chiaverini2004}.
This approach relies on high-fidelity gates for detecting as well as
correcting errors. Another strategy is to passively protect quantum
information by storing the information in a decoherence-free subspace
(DFS)\cite{dfs_intro}. This method has been implemented using
photons\cite{dfs_photons,dfs_photons2}, nuclear magnetic resonance
(NMR) systems\cite{dfs_nmr}, and trapped
ions\cite{LongLifeBell_Roos,RobustEntanglement_Haeffner,dfs_Wineland}.
DFS-encoding has given rise to an impressive increase in coherence
time of more than a factor of one hundred \cite{LongLifeBell_Roos}.
However, the use of DFSs for quantum computational operations so far
was restricted to a single implementation in NMR\cite{grover_dfs_nmr}.
The main challenge is to find methods to implement a universal set of
gates within DFS for a given physical setup and to have sufficient
experimental control to perform these gates with high fidelity.

Typically, ion trap quantum computers rely on quantum bits encoded in
long lived electronic states of individual ions. Nevertheless, the
phase of the qubits can deteriorate quickly which leads to a loss of
encoded information. This dephasing is caused by random fluctuations
of the energy difference between the qubit states $\ket{0}$ and
$\ket{1}$. For qubits based on optical transitions in atoms, this
decoherence mechanism is mainly caused by magnetic field noise and
frequency fluctuations of the laser driving the qubit transition. In
addition, for upcoming realizations of a scalable quantum computer
based on segmented ion traps, ions are moved across sizable distances
where magnetic field gradients lead to an additional, uncontrolled
phase evolution \cite{Kielpinski_scal_qc}. All these types of
dephasing can be overcome by encoding information in a logical qubit,
realized by two physical qubits of the form $\ket{0}_{L}$ =
$\ket{1}_{P} \otimes \ket{0}_{P}\equiv \ket{10}_{P}$ and $\ket{1}_{L}$
= $\ket{0}_{P} \otimes \ket{1}_{P}\equiv\ket{01}_{P}$ where the
indices $P$ and $L$ denote the physical and logical basis,
respectively. Ideally, the energy difference between two logical
states vanishes and energy shifts common to both physical qubits do
not affect the energy difference of the logical eigenstates. Thus the
phase between the two logical qubits is preserved and the logical
eigenstates represent a DFS with respect to the collective decoherence
mechanisms described above.

In this letter, we use this encoding to implement a universal set of
logical gates in a scalable quantum computer. The set is composed of
single-qubit rotations and a two-qubit phase gate acting directly on
the logical qubits. Sequences of such logical gates allow
implementation of arbitrary quantum circuits\cite{arb_qc_1, arb_qc_2}.
As an example, we implement the entangling controlled-NOT (CNOT) gate
operation for logical qubits. We then use the CNOT gate operation,
supplemented with a logical single qubit gate operation, to create
Bell states in the logical subspace. Additionally, we characterize the
CNOT gate operation via quantum process tomography.

\begin{figure}
 \begin{center}
  \includegraphics[width=\columnwidth]{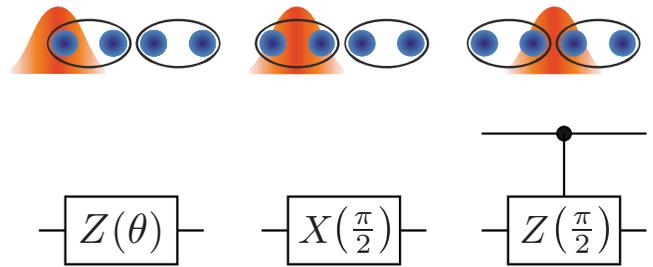}
  \caption{Set of logical gate operations - $\sigma^{z}$ rotation,
    $\sigma^{x}$ rotation and conditional phase gate: A single ion
    AC-Stark shift pulse allows for arbitrary rotation of the logical
    qubit around the z-axis; the MS gate represents a rotation about
    the x-axis of the corresponding Bloch sphere of the logical qubit;
    the phase gate is realized by a copropagating bichromatic light
    field as described in \cite{kiwan_gate}, applied on adjacent ions
    of two logical qubits.}
  \label{fig.set_of_logic_gates}
 \end{center}
\end{figure}

Our experimental system consists of a string of $^{40}$Ca$^+$ ions
trapped in a linear Paul trap. The physical qubits are represented by
the two electronic states
$S_{1/2}(m~=~-1/2)\equiv\ket{S}\equiv\ket{1}_{P}$ and
$D_{5/2}(m=-1/2)\equiv\ket{D}\equiv\ket{0}_{P}$. Individual ion
qubits, or alternatively pairs of ions, are manipulated by a focused
laser beam at 729~nm, exciting the quadrupole transition between the
two states $S_{1/2}$ and $D_{5/2}$ (see Fig.
\ref{fig.set_of_logic_gates}). Optical pumping initializes all ion
qubits in the $\ket{S}$ state, while Doppler cooling and subsequent
sideband cooling prepares the ion string in the motional ground state
of the axial center-of-mass mode. An additional bit flip on one of the
ions initializes the logical qubits in $\ket{0}_{L}=\ket{10}_{P}$ or
$\ket{1}_{L}=\ket{01}_{P}$. Final state detection is performed using
electron shelving of all ions on the $S_{1/2} \leftrightarrow P_{1/2}$
transition by detecting the ions' resonance fluorescence with a CCD
camera. Absence or presence of light corresponds to a projective
measurement in the physical qubit basis $\ket{1}_{P}$ and
$\ket{0}_{P}$, respectively. Details of the setup can be found in Ref.
\cite{expsetup}.

The presented work is largely based on the subsequent application of
two different bichromatic gates, the conditional phase (CP)
gate\cite{kiwan_gate} and the M{\o}lmer and S{\o}rensen (MS)
gate\cite{ms_gate}. For both gates, the qubit-qubit interaction is
mediated by coupling the ions to a motional mode of the ion string.
The gate mechanism can be described by an off-resonantly driven
quantum harmonic oscillator with a state-dependent driving force
\cite{Lee_drivenHO}, i.~e. a Hamiltonian $H \propto (a \exp(i \delta
t) + a^\dagger \exp(-i \delta t)) S_{i}$ where $a^\dagger$ and $a$
represent creation and annihilation operators for motional quanta, and
$\delta$ defines the detuning from the motional sideband. For the
conditional phase gate, the coupling is given by
$S_{i}=\sigma_z^{(1)}+\sigma_z^{(2)}$, whereas in the case of the
M{\o}lmer and S{\o}rensen gate $S_{i}=\sigma_x^{(1)}+\sigma_x^{(2)}$.
After an interaction time $\tau=2\pi/\delta$, the harmonic oscillator
returns to its initial state so that the gate acts only on the
internal states of the ions. The gate action can be described as if
induced by an effective Hamiltonian $H_{\mathrm{eff}}\propto
S_{i}^{2}$ that is nonlinear in the spin operators. In the following,
we present how to rewrite these gates in the basis of logical qubits,
thus representing building blocks for encoded quantum information.

A universal set of gate operations can be realized with arbitrary
single qubit rotations in conjunction with a universal two qubit gate.
In the following we will discuss our experimental realization of such
a set of operations in a DFS as proposed in \cite{AolitaDFS}. First,
we consider arbitrary operations on a single logical qubit. We use
rotations around the z-axis ($\sigma^{z}_{L}$) and the x-axis
($\sigma^{x}_{L}$) of the logical qubit's Bloch sphere. The z-rotation
is implemented by addressing a single physical qubit with a laser beam
far detuned from resonance that shifts the qubits energy levels due to
the AC-Stark effect (see Fig. \ref{fig.set_of_logic_gates}). Because
of the identity $1_{P} \otimes \sigma^{z}_{P} \equiv \sigma^{z}_{L}$,
the z-rotation of one physical qubit directly translates into a
z-rotation on the logical qubit. Rotations by an arbitrary angle
$Z(\theta) \equiv \exp(-i\ \theta/2\ \sigma^{z}_{L})$ are controlled
by the intensity and pulse length of the laser pulse. Fidelities of
98(1)\% were measured for the $\sigma^{z}_{L}$ gate using Ramsey
experiments.

The second single logical qubit gate, the $\sigma^{x}_{L}$ rotation,
requires collective operation on both physical qubits, since
$\sigma^{x}_{L} \equiv \sigma^{x}_{P} \otimes \sigma^{x}_{P}$. To
realize this operation, a focused beam is centered between a pair of
ions, equally illuminating both with a bichromatic light field as
described by M{\o}lmer and S{\o}rensen \cite{ms_gate}. The two
frequencies are $\omega_{0} \pm (\omega_{z} + \delta_\mathrm{MS})$, where
$\omega_{0}$ is the transition frequency from $|S\rangle$ to
$|D\rangle$, $\omega_{z}$ represents the frequency of the
center-of-mass mode in the axial direction ($\omega_z\approx (2\pi)$
1.2~MHz) and $\delta_\mathrm{MS}$ is a detuning set to $(2\pi)$ 7~kHz. The
light field intensity is chosen such that this operation rotates the
corresponding Bloch sphere of the single logical qubit by $\pi/2$ (in
the following referred to as $X(\pi/2) \equiv \exp(-i\ \pi/4\
\sigma^{x}_{L}$)) after a gate time $\tau$ of
$\tau(X(\pi/2))=2\pi/\delta_\mathrm{MS} = \mathrm{143}\mu s$. Our
characterization of this operation by measuring the final populations
combined with parity oscillations on the output state\cite{didi_paper}
indicates a fidelity for a logical $\pi/2$ pulse of 96(2)\%.

The set of universal gates is completed by a two-qubit interaction,
here a conditional phase gate. For operations on two logical qubits we
work on a string of four ions, where ions 1 and 2 (3 and 4) represent
the first (second) logical qubit. Performing a phase gate on the
center physical qubits (2 and 3) will translate into a phase gate
$\sigma^{z}_{L} \otimes \sigma^{z}_{L}$ acting on the logical qubits
(also see Ref. \cite{AolitaDFS}). For this purpose, the two center
ions are illuminated by a bichromatic focused beam realizing a
$\sigma^{z}_{P} \otimes \sigma^{z}_{P}$ gate as proposed in
\cite{kiwan_gate}. The frequencies of the bichromatic laser beam are
set to $\omega_{0} \pm 1/2 (\omega_{z} + \delta_\mathrm{CP})$. For our
implementation a detuning of about 2~kHz was chosen for $\delta_\mathrm{CP}$,
resulting in a gate time of $\tau(CP)=2\pi/\delta_\mathrm{CP} =
\mathrm{470}\mu s$. The performance of this novel phase gate has been
investigated by applying it to two ions and carrying out process
tomography\cite{proctomo}, obtaining a mean gate fidelity of 94(1)\%.

\begin{figure}
 \begin{center}
  \includegraphics[width=\columnwidth]{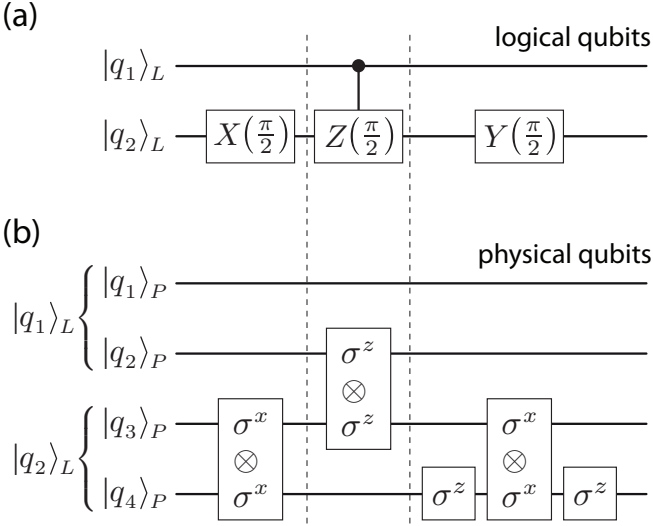}
  \caption{Pulse sequence to realize a controlled-NOT operation within
    a DFS - The logical pulse sequence (a) and the operations on the
    physical qubits (b) are depicted: A logical phase gate is
    performed on two logical qubits. For the target qubit, the phase
    gate is enclosed by two Ramsey pulses, respectively $\pi/2$
    rotations along the x- and y- axis of the corresponding Bloch
    sphere (represented by a composite
    $\sigma^{z}\sigma^{x}\sigma^{z}$ rotation).}
  \label{fig.pulse_sequence}
 \end{center}
\end{figure}

We combine single- and two-qubit logical gates to implement the
entangling CNOT operation within the chosen DFS. To this end, the
logical phase gate is enclosed by two Ramsey pulses on the logical
target qubit (see Fig. \ref{fig.pulse_sequence}). Depending on the
control state the second Ramsey pulse will either flip the target
qubit or recover its initial state. Experimentally, we achieve an
improved gate fidelity by splitting the phase gate into two pulses,
allowing for a spin echo pulse on both logical qubits. The resulting
ideal unitary matrix associated with our CNOT operation is:
\[ U_{\textrm{CNOT}}=
\left( \begin{array}{cccc}
    0 &  -1 &   0 &   0 \\
    i &   0 &   0 &   0 \\
    0 &   0 &   1 &   0 \\
    0 &   0 &   0 &   i
\label{matrix:unitary_map_CNOT}
  \end{array} \right)\]

\begin{figure}
 \begin{center}
  \includegraphics[width=\columnwidth]{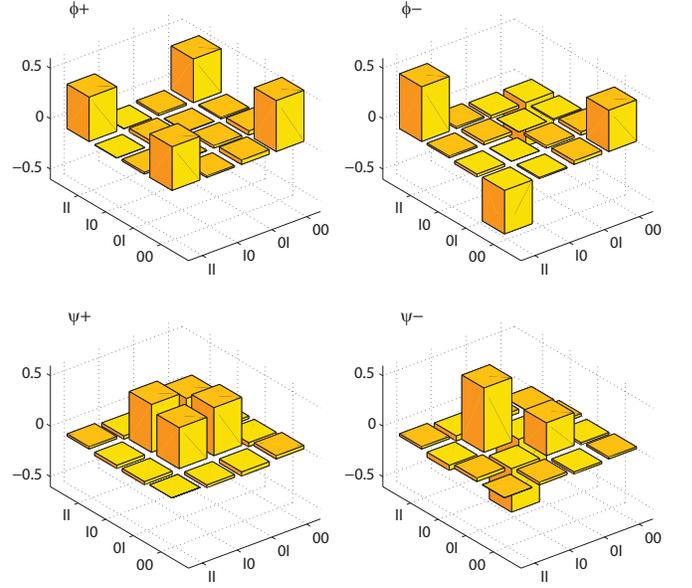}
  \caption{The DFS-CNOT gate can be used to generate Bell states in
    the logical qubit Hilbert space. For a given superposition between
    $\ket{0}_{L}$ and $\ket{1}_{L}$ on the control qubit, the
    resulting output state will be one of the four Bell states. The
    real part of the obtained density matrices for all four Bell
    states is shown above. On average, a fidelity of 91(2)\% is
    achieved within the decoherence-free subspace.}
  \label{fig.dfs_bellstates}
 \end{center}
\end{figure}

In order to prove that the gate acts as intended, this CNOT gate is
applied to generate entangled states within the DFS. After preparation
of one of the four basis states
\{$\ket{00}_{L},\ket{01}_{L},\ket{10}_{L},\ket{11}_{L}$\}, a
superposition between the logical ground and excited state of the
control qubit is generated by a $X(\pi/2)$ rotation. Subsequent
application of the CNOT gate directly maps the input states onto the
Bell basis states
\{$\ket{\phi^{+}}_{L},\ket{\phi^{-}}_{L},\ket{\psi^{+}}_{L},\ket{\psi^{-}}_{L}$\}
defined by $\ket{\phi^{\pm}}_{L}=\ket{00}_{L}\pm\ket{11}_{L}$ and
$\ket{\psi^{\pm}}_{L}=\ket{01}_{L}\pm\ket{10}_{L}$. In the physical
basis, these states are equivalent to four-qubit Schr\"odinger cat
states. The output state is determined by quantum state tomography in
the Hilbert space of the four physical qubits. Restricting quantum
computation to a subspace of the total Hilbert space leads to two
basic questions at the end of a computation: a) Is the outcome within
the subspace? b) How close is the result to the expected one?
Accepting only results within the DFS allows for computation at higher
fidelities, but making it probabilistic. The probability of a
state to remain in the DFS after application of a certain gate
sequence will be called permanence $P$. The fidelity of the generated
state within the DFS can be calculated in a straightforward manner.
The four Bell basis states are obtained with fidelities $F$ of
\{89(1),91(1),91(1),92(1)\}\% and a permanence $P$ of
\{90.2,94.3,83.9,86.0\}\%. The real part of the obtained density
matrices of the four different Bell states are depicted in
Fig.~\ref{fig.dfs_bellstates}.

In order to fully characterize the logical CNOT gate, its performance
is analyzed by quantum process tomography \cite{proctomo}. For the
given two-logical qubit Hilbert space it is performed by creating
$4^2=16$ linear independent logical input states, applying the CNOT
gate, and fully characterizing the output state via state tomography
in the physical basis ($3^4$ settings each). Each experimental setting
was repeatedly measured 100 times and averaged, resulting in a total
measurement time per setting of about 5 seconds. In total the
characterization of the CNOT gate requires about 2~hours of
measurement time. Evaluation of this data allows us to derive the
so-called $\chi$-matrix that describes the investigated process
$\mathcal{E}$, here the CNOT within a DFS, such that
$\rho_{out}=\mathcal{E}(\rho_{in})=\sum_{m,n=1}^{4^N} \chi_{m,n} A_{m}
\rho_{in} A^{\dagger}_{n}$, where $N$ is the number of logical qubits,
$\rho_{in}$ and $\rho_{out}$ are the input and output density matrices
and $A$ is a basis of operators in the Hilbert space of dimension $2^N
\times 2^N$ \cite{chuangblackbox}. Taking $2 \times 10^{5}$ pure
logical input states, randomly drawn from the unitary group $U(4)$
according to the Haar measure \cite{Pozniak1998JoPAMaGv31p1059--1071},
the mean gate fidelity is then calculated by
$\bar{F}=\mathrm{mean}_{\psi_{i}}\left[\langle \psi_{i}| U^{\dagger}\
  \mathcal{E}(|\psi_{i}\rangle \langle \psi_{i}|)\ U | \psi_{i}
  \rangle\right]$, where $U$ represents the ideal unitary map for the
implemented process. We infer a mean permanence of the CNOT operation
of $\bar{P}=$ 89(7)\% and a mean gate fidelity of $\bar{F}=$ 89(4)\%
within the DFS. The overall fidelity of $\bar{P} \cdot \bar{F} \approx
79(7)\%$ is consistent with the achieved fidelities of its constituent
operations of about $83(3)\%$. The mean gate fidelity within the DFS
of $89(4)\%$ is comparable with current state-of-the-art two-qubit
quantum gates acting on selected qubits out of a quantum register,
operating at a fidelity of 92.6(6)\% \cite{proctomo}.


Infidelities of the gate can be classified according to their effect
on the permanence or gate fidelity within the DFS. Addressing errors
constitute the main error source for leaving the DFS during a pulse
sequence. When focusing the laser down onto a pair of two ions, some
residual light is also applied to the adjacent ions. The addressing
error is characterized by the ratio of the Rabi frequency of a logical
qubit compared to the Rabi frequency of an adjacent, single ion. For
the chosen parameters the ratio was about 5\%. This unwanted
excitation on the neighboring ion results in a population loss from
the DFS. Another error source for leaving the DFS are off-resonant
excitations during the bichromatic gates. To minimize this error we
use amplitude-shaped laser pulses as described in
\cite{amplitude_shaping}. Errors within the DFS are mainly due to
unbalanced intensities of the light fields acting on the two
simultaneously addressed ions. The difference between the Rabi
frequencies is caused by beam pointing instability with regard to the
ion position. Finally, intensity fluctuations lead directly to phase
errors of the single-logical qubit phase gate via the intensity
dependence of the AC-Stark effect. Note that laser frequency and
magnetic field fluctuations do not contribute to errors since the DFS
encoding protects against such decoherence. All shortcomings described
above are caused by technical imperfections and do not represent a
fundamental limit to the achievable fidelities. Spontaneous decay of
the excited level as the only fundamental error can be avoided by
encoding the physical qubits in the two Zeeman-ground states.

To conclude, we have demonstrated a universal set of quantum gates
acting in a decoherence-free subspace of trapped ions, consisting of
addressable gate operations, namely: single logical qubit
$\sigma^{z}_{L}$ and $\sigma^{x}_{L}$ as well as a two logical qubit
phase gate $\sigma^{z}_{L} \otimes \sigma^{z}_{L}$. Using these gates
we have implemented and characterized a controlled-NOT gate within a
decoherence free subspace acting on logical ion qubits. Our
implementation achieves fidelities close to current state-of-the-art
quantum computation as well as it employs logical qubits with a
coherence time one hundred times longer than their single
constituents.

\begin{acknowledgments}
  We gratefully acknowledge support by the Austrian Science Fund
  (FWF), by the European Commission (SCALA), and by the Institut f\"ur
  Quanteninformation GmbH. This material is based upon work supported
  in part by IARPA. T.M. and K.K. contributed equally to this work.
\end{acknowledgments}


\begin{thebibliography}{24}
\expandafter\ifx\csname natexlab\endcsname\relax\def\natexlab#1{#1}\fi
\expandafter\ifx\csname bibnamefont\endcsname\relax
  \def\bibnamefont#1{#1}\fi
\expandafter\ifx\csname bibfnamefont\endcsname\relax
  \def\bibfnamefont#1{#1}\fi
\expandafter\ifx\csname citenamefont\endcsname\relax
  \def\citenamefont#1{#1}\fi
\expandafter\ifx\csname url\endcsname\relax
  \def\url#1{\texttt{#1}}\fi
\expandafter\ifx\csname urlprefix\endcsname\relax\def\urlprefix{URL }\fi
\providecommand{\bibinfo}[2]{#2}
\providecommand{\eprint}[2][]{\url{#2}}

\bibitem[{\citenamefont{Steane}(1996)}]{Steane_ErrorCorrection}
\bibinfo{author}{\bibfnamefont{A.}~\bibnamefont{Steane}},
  \bibinfo{journal}{Proc. Roy. Soc. A} \textbf{\bibinfo{volume}{452}},
  \bibinfo{pages}{2551} (\bibinfo{year}{1996}).

\bibitem[{\citenamefont{Calderbank and
  Shor}(1996)}]{Calderbank_ErrorCorrection}
\bibinfo{author}{\bibfnamefont{A.~R.} \bibnamefont{Calderbank}}
  \bibnamefont{and} \bibinfo{author}{\bibfnamefont{P.~W.} \bibnamefont{Shor}},
  \bibinfo{journal}{Phys. Rev. A} \textbf{\bibinfo{volume}{54}},
  \bibinfo{pages}{1098} (\bibinfo{year}{1996}).

\bibitem[{\citenamefont{Chiaverini et~al.}(2004)}]{Chiaverini2004}
\bibinfo{author}{\bibfnamefont{J.}~\bibnamefont{Chiaverini}},
  \bibnamefont{et~al.}, \bibinfo{journal}{Nature}
  \textbf{\bibinfo{volume}{432}}, \bibinfo{pages}{602} (\bibinfo{year}{2004}).

\bibitem[{\citenamefont{Lidar et~al.}(1998)\citenamefont{Lidar, Chuang, and
  Whaley}}]{dfs_intro}
\bibinfo{author}{\bibfnamefont{D.~A.} \bibnamefont{Lidar}},
  \bibinfo{author}{\bibfnamefont{I.~L.} \bibnamefont{Chuang}},
  \bibnamefont{and} \bibinfo{author}{\bibfnamefont{K.~B.}
  \bibnamefont{Whaley}}, \bibinfo{journal}{Phys. Rev. Lett.}
  \textbf{\bibinfo{volume}{81}}, \bibinfo{pages}{2594} (\bibinfo{year}{1998}).

\bibitem[{\citenamefont{Kwiat et~al.}(2000)\citenamefont{Kwiat, Berglund,
  Altepeter, and White}}]{dfs_photons}
\bibinfo{author}{\bibfnamefont{P.}~\bibnamefont{Kwiat}},
  \bibnamefont{et~al.},
  \bibinfo{journal}{Science} \textbf{\bibinfo{volume}{290}},
  \bibinfo{pages}{498} (\bibinfo{year}{2000}).

\bibitem[{\citenamefont{Bourennane et~al.}(2004)\citenamefont{Bourennane, Eibl,
  G\"artner, Kurtsiefer, Cabello, and Weinfurter}}]{dfs_photons2}
\bibinfo{author}{\bibfnamefont{M.}~\bibnamefont{Bourennane}},
  \bibnamefont{et~al.},
  \bibinfo{journal}{Phys. Rev. Lett.} \textbf{\bibinfo{volume}{92}},
  \bibinfo{pages}{107901} (\bibinfo{year}{2004}).

\bibitem[{\citenamefont{Viola et~al.}(2001)\citenamefont{Viola, Fortunato,
  Pravia, Knill, Laflamme, and Cory}}]{dfs_nmr}
\bibinfo{author}{\bibfnamefont{L.}~\bibnamefont{Viola}},
  \bibnamefont{et~al.},
  \bibinfo{journal}{Science} \textbf{\bibinfo{volume}{293}},
  \bibinfo{pages}{2059 } (\bibinfo{year}{2001}).

\bibitem[{\citenamefont{Roos et~al.}(2004)\citenamefont{Roos, Lancaster, Riebe,
  H\"affner, H\"ansel, Gulde, Becher, Eschner, Schmidt-Kaler, and
  Blatt}}]{LongLifeBell_Roos}
\bibinfo{author}{\bibfnamefont{C.~F.} \bibnamefont{Roos}},
  \bibnamefont{et~al.},
  \bibinfo{journal}{Phys. Rev. Lett.} \textbf{\bibinfo{volume}{92}},
  \bibinfo{pages}{220402} (\bibinfo{year}{2004}).

\bibitem[{\citenamefont{H\"affner et~al.}(2005)\citenamefont{H\"affner,
  Schmidt-Kaler, H\"ansel, Roos, Körber, Chwalla, Riebe, Benhelm, Rapol,
  Becher et~al.}}]{RobustEntanglement_Haeffner}
\bibinfo{author}{\bibfnamefont{H.}~\bibnamefont{H\"affner}},
  \bibnamefont{et~al.},
  \bibnamefont{et~al.}, \bibinfo{journal}{Appl. Phys. B}
  \textbf{\bibinfo{volume}{81}}, \bibinfo{pages}{151} (\bibinfo{year}{2005}).

\bibitem[{\citenamefont{Kielpinski et~al.}(2001)\citenamefont{Kielpinski,
  Meyer, Rowe, Sackett, Itano, Monroe, and Wineland}}]{dfs_Wineland}
\bibinfo{author}{\bibfnamefont{D.}~\bibnamefont{Kielpinski}},
  \bibnamefont{et~al.},
  \bibinfo{journal}{Science} \textbf{\bibinfo{volume}{291}},
  \bibinfo{pages}{1013} (\bibinfo{year}{2001}).

\bibitem[{\citenamefont{Ollerenshaw et~al.}(2003)\citenamefont{Ollerenshaw,
  Lidar, and Kay}}]{grover_dfs_nmr}
\bibinfo{author}{\bibfnamefont{J.~E.} \bibnamefont{Ollerenshaw}},
  \bibinfo{author}{\bibfnamefont{D.~A.} \bibnamefont{Lidar}}, \bibnamefont{and}
  \bibinfo{author}{\bibfnamefont{L.~E.} \bibnamefont{Kay}},
  \bibinfo{journal}{Phys. Rev. Lett.} \textbf{\bibinfo{volume}{91}},
  \bibinfo{pages}{217904} (\bibinfo{year}{2003}).

\bibitem[{\citenamefont{Kielpinski et~al.}(2002)\citenamefont{Kielpinski,
  Monroe, and Wineland}}]{Kielpinski_scal_qc}
\bibinfo{author}{\bibfnamefont{D.}~\bibnamefont{Kielpinski}},
  \bibinfo{author}{\bibfnamefont{C.}~\bibnamefont{Monroe}}, \bibnamefont{and}
  \bibinfo{author}{\bibfnamefont{D.~J.} \bibnamefont{Wineland}},
  \bibinfo{journal}{Nature} \textbf{\bibinfo{volume}{417}},
  \bibinfo{pages}{709} (\bibinfo{year}{2002}).

\bibitem[{\citenamefont{Sleator and Weinfurter}(1995)}]{arb_qc_1}
\bibinfo{author}{\bibfnamefont{T.}~\bibnamefont{Sleator}} \bibnamefont{and}
  \bibinfo{author}{\bibfnamefont{H.}~\bibnamefont{Weinfurter}},
  \bibinfo{journal}{Phys. Rev. Lett.} \textbf{\bibinfo{volume}{74}},
  \bibinfo{pages}{4087} (\bibinfo{year}{1995}).

\bibitem[{\citenamefont{DiVincenzo}(1995)}]{arb_qc_2}
\bibinfo{author}{\bibfnamefont{D.~P.} \bibnamefont{DiVincenzo}},
  \bibinfo{journal}{Phys. Rev. A} \textbf{\bibinfo{volume}{51}},
  \bibinfo{pages}{1015} (\bibinfo{year}{1995}).

\bibitem[{\citenamefont{Schmidt-Kaler et~al.}(2003)\citenamefont{Schmidt-Kaler,
  H\"affner, Gulde, Riebe, Lancaster, Deuschle, Becher, Hansel, Eschner, Roos
  et~al.}}]{expsetup}
\bibinfo{author}{\bibfnamefont{F.}~\bibnamefont{Schmidt-Kaler}},
  \bibnamefont{et~al.},
 \bibinfo{journal}{Appl. Phys. B}
  \textbf{\bibinfo{volume}{77}}, \bibinfo{pages}{789} (\bibinfo{year}{2003}).

\bibitem[{\citenamefont{Kim et~al.}(2008)\citenamefont{Kim, Roos, Aolita,
  H\"affner, Nebendahl, and Blatt}}]{kiwan_gate}
\bibinfo{author}{\bibfnamefont{K.}~\bibnamefont{Kim}},
  \bibnamefont{et~al.},
  \bibinfo{journal}{Phys. Rev. A} \textbf{\bibinfo{volume}{77}},
  \bibinfo{pages}{050303(R)} (\bibinfo{year}{2008}).

\bibitem[{\citenamefont{S{\o}rensen and M{\o}lmer}(1999)}]{ms_gate}
\bibinfo{author}{\bibfnamefont{A.}~\bibnamefont{S{\o}rensen}} \bibnamefont{and}
  \bibinfo{author}{\bibfnamefont{K.}~\bibnamefont{M{\o}lmer}},
  \bibinfo{journal}{Phys. Rev. Lett.} \textbf{\bibinfo{volume}{82}},
  \bibinfo{pages}{1971} (\bibinfo{year}{1999}).

\bibitem[{\citenamefont{Lee et~al.}(2005)\citenamefont{Lee, Brickman,
  Deslauriers, Haljan, Duan, and Monroe}}]{Lee_drivenHO}
\bibinfo{author}{\bibfnamefont{P.}~\bibnamefont{Lee}},
  \bibnamefont{et~al.},
  \bibinfo{journal}{J. Opt. B} \textbf{\bibinfo{volume}{7}},
  \bibinfo{pages}{S371} (\bibinfo{year}{2005}).

\bibitem[{\citenamefont{Aolita et~al.}(2007)\citenamefont{Aolita, Davidovich,
  Kim, and H\"affner}}]{AolitaDFS}
\bibinfo{author}{\bibfnamefont{L.}~\bibnamefont{Aolita}},
  \bibnamefont{et~al.},
  \bibinfo{journal}{Phys. Rev. A} \textbf{\bibinfo{volume}{75}},
  \bibinfo{eid}{052337} (\bibinfo{year}{2007}).

\bibitem[{\citenamefont{Leibfried et~al.}(2003)\citenamefont{Leibfried,
  Demarco, Meyer, Lucas, Barrett, Britton, Itano, Jelenkovic, Langer, Rosenband
  et~al.}}]{didi_paper}
\bibinfo{author}{\bibfnamefont{D.}~\bibnamefont{Leibfried}},
  \bibnamefont{et~al.},
 \bibinfo{journal}{Nature}
  \textbf{\bibinfo{volume}{422}}, \bibinfo{pages}{412} (\bibinfo{year}{2003}).

\bibitem[{\citenamefont{Riebe et~al.}(2006)\citenamefont{Riebe, Kim, Schindler,
  Monz, Schmidt, K\"orber, H\"ansel, H\"affner, Roos, and Blatt}}]{proctomo}
\bibinfo{author}{\bibfnamefont{M.}~\bibnamefont{Riebe}},
  \bibnamefont{et~al.},
  \bibinfo{journal}{Phys. Rev. Lett.} \textbf{\bibinfo{volume}{97}},
  \bibinfo{pages}{220407} (\bibinfo{year}{2006}).


\bibitem[{\citenamefont{Chuang and Nielsen}(1997)}]{chuangblackbox}
\bibinfo{author}{\bibfnamefont{I.}~\bibnamefont{Chuang}} \bibnamefont{and}
  \bibinfo{author}{\bibfnamefont{M.}~\bibnamefont{Nielsen}},
  \bibinfo{journal}{J. Mod. Opt.} \textbf{\bibinfo{volume}{44}},
  \bibinfo{pages}{2455} (\bibinfo{year}{1997}).

\bibitem[{\citenamefont{Pozniak et~al.}(1998)\citenamefont{Pozniak, Zyczkowski,
  and Kus}}]{Pozniak1998JoPAMaGv31p1059--1071}
\bibinfo{author}{\bibfnamefont{M.}~\bibnamefont{Pozniak}},
  \bibinfo{author}{\bibfnamefont{K.}~\bibnamefont{Zyczkowski}},
  \bibnamefont{and} \bibinfo{author}{\bibfnamefont{M.}~\bibnamefont{Kus}},
  \bibinfo{journal}{J. Phys. A: Math. Gen.}
  \textbf{\bibinfo{volume}{31}}, \bibinfo{pages}{1059} (\bibinfo{year}{1998}).

\bibitem[{\citenamefont{Roos}(2008)}]{amplitude_shaping}
\bibinfo{author}{\bibfnamefont{C.~F.} \bibnamefont{Roos}},
  \bibinfo{journal}{New J. Phys.} \textbf{\bibinfo{volume}{10}},
  \bibinfo{pages}{013002} (\bibinfo{year}{2008}).

\end{thebibliography}

\end{document}